\documentclass[12pt]{article}
 \usepackage[dvips]{graphicx}
\begin{document}
\title { Schroedinger revisited:How the time-dependent wave equation follows from
the Hamilton-Jacobi equation }
\author{Alex Granik\thanks{Department of
Physics,UOP,Stockton,CA.95211;agranik@pacific.edu;}}
\bigskip
\date{}
\maketitle
\begin{abstract}

It is shown how using the classical Hamilton-Jacobi equation one
can arrive at the time-dependent wave equation. Although the
former equation was originally used by E.Schroedinger to get the
wave equation, we propose a different approach. In the first
place, we do not use the principle of least action and, in
addition, we arrive at the time-dependent equation, while
Schroedinger (in his first seminal paper) used the least action
principle and obtained the stationary wave equation. The proposed
approach works for any classical Hamilton-Jacobi equation. In
addition, by introducing information loss into the Hamilton-Jacobi
equation we derive in an elementary fashion the wave equations
(ranging from the Shroedinger to Klein-Gordon, to Dirac
equations). We also apply this technique to a relativistic
particle in the gravitational field and obtain the respective wave
equation. All this supports 't Hooft's proposal about a
possibility of arriving at quantum description from a classical
continuum in the presence of information loss.
\end{abstract}

\section{Introduction}In a first of his historical papers \cite{S1}
E.Schroedinger looking for a solution to the quantization problem
of energy values for the hydrogen atom applied the variational
principle to the $stationary$ classical Hamilton-Jacobi equation.
He replaced the classical action $S$ with a new function $\Psi$
and obtained the $\it stationary$ Schroedinger equation. In a
sense, it was an {\it ad hoc} approach, since quoting R.Feynman,
this equation "is impossible to derive from anything. It came out
of the mind of Schroedinger". \\

Our paper consists of 2 sections. In the first section we revisit
the above Schroedinger approach, but take a different path, which
does not require the variational principle and results (with the
help of elementary calculations)in the $time-dependent$ wave
equation. By contrast, the latter equation was obtained by
Schroedinger "by way of trial" \cite{ES3}, which was used to
resolve the bi-harmonic wave equations into two  partial
differential equations. Matching these 2 equations, he arrived at
the wave equation.\\

The second section deals with the recent proposal of t' Hooft
\cite{GH}which argues that introduction of information loss (in a
form of small perturbations) into certain classical systems may
lead "an apparent quantization of the orbits which resembles the
quantum structure seen in the real world". This proposal was
echoed in \cite{AG2} where it was pointed out that "Whereas the
entropy of any deterministic classical system described by a
principle of least action is zero, one can assign a "quantum
information" to quantum mechanical degree of freedom
equal to Hausdorff area of the deviation from a classical path."\\

Here we show how in a $more ~general$ scheme of things (not
restricted to some special cases) that adding a dissipative term
to the classical equation of motion will naturally lead to the
wave equations, ranging from the Shroedinger to Klein-Gordon, to
Dirac equations. We also apply this method to the motion of a
particle in the gravitational field and obtain the respective wave
equation. All this makes the concept of information loss
introduction into a classical system as a very promising one.\\

Interestingly enough, such an idea, in a sense, represents an
inverse of the "hidden variable" approach to quantum mechanics by
Bohm \cite{B}. Whereas Bohm considered the quantum potential
arising directly from the Schroedinger equation, the information
loss idea takes as its starting point the equations of classical
mechanics.

\section{Elementary derivation of the Schroedinger equation}
We begin with the classical Hamilton-Jacobi equation for a single
particle of a mass $m$ moving in a constant potential field $U$
\begin{equation}
\label{1} \frac{\partial{S}}{\partial t}+\frac{1}{2m}(\nabla
S)^2+U=0 \end{equation}

Equation (\ref{1}) is a nonlinear inhomogeneous PDE with constant
coefficients. On the other hand, a nonlinear homogeneous
differential equation with constant coefficients has oscillatory
solutions. Thus if it would be possible to transform the
Hamilton-Jacobi equation ( by a suitable choice of a new function
instead of action $S$) to the respective nonlinear (in this case
quadratic) homogeneous differential equation, it would yield the
oscillatory solution. This would unify the particle and wave
descriptions implicit in the Hamilton-Jacobi equation. \

Let us introduce a new variable, say $\Psi(\vec{x},t)$:
\begin{equation}
\label{2} S=S[\Psi(\vec{x},t)]
\end{equation}
As a result, the Hamilton-Jacobi equation (\ref{1}) is reduced to
the following equation
\begin{equation}
\label{3} \frac{1}{\partial S/\partial\Psi}\frac{\partial
\Psi}{\partial t}+\frac{\hbar^2}{2m}(\nabla
\Psi)^2+U\frac{1}{(\partial S/\partial \Psi)^2}=0
\end{equation}
It is clear that this equation becomes a nonlinear ( quadratic)
homogeneous differential equation if and only if
\begin{equation}
\label{4} \frac{\partial S}{\partial\Psi}=
\frac{B}{\Psi}\longrightarrow\space S=BLn(A\Psi)
\end{equation}
where $A$ is a constant of integration and $B$ is a constant to be
found. Without any loss of generality we can take $A=1$.  On the
basis of the dimensional considerations ( and to make what follows
compatible with the well-known results) we represent constant $B$
as
\begin{equation}
\label{add}
B=a\hbar
\end{equation}
where $\hbar$ is the Planck constant and constant $a$ will be
found in what follows.

Hence Eq.(\ref{1}) becomes
\begin{equation}
\label{5} a\hbar\Psi\frac{\partial\Psi}{\partial
t}+\frac{a^2\hbar^2}{2m}(\nabla\Psi)^2+U\Psi^2=0
\end{equation}
If $U=const$ then (\ref{5}) admits a wave solution
\begin{equation}
\label{6} \Psi=\alpha e^{i(\gamma t-\vec{\beta}\cdot\vec{x})}
\end{equation}

Since the linear momentum is $\vec{p}=\nabla S$ and the energy is
$E=-\partial S/\partial t$ we obtain from (\ref{6}) the values of
constants $\gamma$ and $\vec{\beta}$
\begin{equation}
\label{7} \gamma=\frac{i}{a\hbar}E,~~
\vec{\beta}=\frac{i}{a\hbar}\vec{p}
\end{equation}
From (\ref{7}) follows that to preserve the wave character of the
solution (\ref{6}) it is necessary for $a$ to be an imaginary
quantity. To keep the notation accepted in the literature, we set
(without any loss of generality) $$a=\frac{1}{i}$$ yielding
according to (\ref{4}) and (\ref{add})
\begin{equation}
\label{a1} S=\frac{\hbar}{i}Ln\Psi
\end{equation}
Therefore now  action $S$ becomes a $complex-valued$ quantity.\\

The change of variable (\ref{a1}) requires a more $rigorous~ and~
physically~ justifiable$ argumentation. Therefore we provide
additional reasons based on a dual (particle-like and wave-like)
nature of the Hamilton-Jacobi equation which, so-to-speak, offers
more than sees an eye. To this end we consider the relativistic
Hamilton-Jacobi equation for a massless particle (in itself a
rather strange, but still valid, concept within the framework of
classical mechanics)
\begin{equation}
\label{al1} \frac{\partial^2S}{\partial t^2}=(\nabla S)^2
\end{equation}
where we set $c=1$.\

One can easily see that it has two different solutions. One, let's
call it particle-like, is
\begin{equation}
\label{al2} S_p=-Et+\vec{p}\cdot\vec{x}
\end{equation}
Another one, let's call it  wave-like, is
\begin{equation}
\label{al3} S_w=exp[-i(\omega t-\vec{k}\cdot\vec{x})]
\end{equation}
On one hand, from (\ref{al2})
 \begin{equation} \label{new} \frac{\partial
S_p}{\partial t}=-E
\end{equation}
 and from (\ref{al3})
\begin{equation}
\label{newa} \frac{\partial}{\partial t}(\frac{1}{i}Ln
S_w)=-\omega
\end{equation}\

On the other hand, according to Planck's hypothesis about a
discrete character of energy transfer, we replace in (\ref{new})
(for a single massless particle) energy
 $E$ by $\hbar\omega$, which yields
$$\frac{\partial }{\partial t}(\frac{S_p}{\hbar})= -\omega$$
Comparing this with the right hand-side  of (\ref{newa}) we obtain
the unique relation between two solutions, $S_p$ and $S_w$:
\begin{equation}
\label{a5} S_p=\frac{\hbar}{i}Ln S_w\equiv\frac{\hbar}{i}Ln\Psi
\end{equation}
Now comparing $$\nabla S_p=\vec{p}~~~~ and
~~~~\nabla(\frac{1}{i}Ln S_w)=\vec{k}$$ we find from (\ref{a5})
the De Broglie formula
\begin{equation}
\label{al1}
 \vec{p}=\hbar\vec{k}
 \end{equation}
 Thus the dual character ( wave-like and particle-like) of a
solution to the Hamilton-Jacobi equation inevitably leads to the
emergence of the $complex-valued$ wave
 "action" $S_w$ (wave function $\Psi$) related to the particle
action $S_p$ via a naturally arising substitution (\ref{a5})which
Schroedinger originally introduced "by hand" \cite{S1}.\\

This holds true for the case $U=U(\vec{x},t)$, where we have to
modify the expression for $\Psi$ (Eq.\ref{6}) by requiring the
quantity $\alpha$ to become a function of $\vec{x}$ and $t$
$$\alpha=d(\vec{x},t)$$ and replacing the power in the exponent by
some real-valued function $f(\vec{x},t)$:
\begin{equation}
\label{8} \Psi=d(\vec{x},t)e^{if(\vec{x},t)/\hbar}
\end{equation}
This means that $$S=\frac{\hbar}{i}Lnd+f(\vec{x},t)$$\\

It is clear that for
\begin{equation}
\label{a2} |\hbar Lnd|\ll |f|
\end{equation}
the wave part of the complex-valued action (which replaced the
classical action) becomes negligible as compared to the "particle"
part of the action, and we return to the classical particle
description via
classical action $S\longrightarrow f=S_{cl}$.\\

With the help of the dimensional analysis, this transition can be
interpreted as follows. Since now $|p/\hbar|=k$ (where
$k=2\pi/\lambda$ is the wave number and $p\sim m_cL_c/T_c$),
relation (\ref{al1}) allows us to evaluate the quantum of action
$\hbar$ in terms of the characteristic mass $m_c$, length interval
$L_c$, time interval $T_c$ and the wavelength $\lambda$ entering
into a description of the wave:
\begin{equation}
\label{a3} |\frac{2\pi}{\lambda}|=|\frac{p}{\hbar}|\sim\frac{2\pi
m_cL_c}{hT_c}\longrightarrow h\sim\frac{m_cL_c}{T_c}\lambda
\end{equation}
Substituting (\ref{a3})  into (\ref{a2}) we find the condition
under which the wave-like description is replaced by the
particle-like description
\begin{equation}
\label{a4} \lambda\ll L_c
\end{equation}
where we use $S_c\sim m_cL_c^2/T_c$. This gives us the condition
defining  geometrical optics which is more physically justifiable
than the usual requirement of setting
$\lambda\longrightarrow 0$.\\

Since now the modified action $S$ is {\bf complex-valued}, it
provides a clue how we should treat the term $(\nabla S)^2$ in
(\ref{1}) under such condition. By requiring this term to be
real-valued (which is dictated by the fact that this term in
classical mechanics has a meaning of the kinetic energy) we
perform the following replacement:
 $$(\nabla S)^2\longrightarrow \nabla S\cdot\nabla S^*$$
where $S^*$ is the complex-conjugate of $S$. As a result, with the
help of (\ref{a1}) (where $a=1/i$)
 equation (\ref{5}) for $\Psi$ becomes
 \begin{equation}
 \label{9}
 \frac{\hbar}{i}\Psi^*\frac{\partial\Psi}{\partial
 t}+\frac{\hbar^2}{2m}\nabla\Psi\cdot\nabla\Psi^*+U\Psi\Psi^*=0
 \end{equation}
 The complex-conjugation of (\ref{9}) is respectively
\begin{equation}
 \label{10}
 -\frac{\hbar}{i}\Psi^*\frac{\partial\Psi}{\partial
 t}+\frac{\hbar^2}{2m}\nabla\Psi\cdot\nabla\Psi^*+U\Psi\Psi^*=0
 \end{equation}
By combining (\ref{9}) and (\ref{10}) we obtain a restriction
imposed on the absolute value of the function $\Psi$
\begin{equation}
\label{11}
\frac{\partial}{\partial t}|\Psi|^2=0
\end{equation}
Now it is easy to obtain the equation for $\Psi$ (or $\Psi^*$). To
this end we integrate (\ref{9})(or eq.\ref{10}) over an arbitrary
volume $\tau$:
\begin{equation}
\label{12}
\frac{\hbar}{i}\int\Psi^*\frac{\partial\Psi}{\partial t}d\tau+\frac{\hbar^2}{2m}\int\nabla\Psi^*
\nabla\Psi d\tau+\int\Psi^*\Psi d\tau=0
\end{equation}
Integrating the second term by parts and assuming  that at the
boundary surface $\Sigma$ of the volume $\tau$
$$\Psi|_{\Sigma}=0$$
we obtain from (\ref{12})
\begin{equation}
\label{13}
\int\Psi^*\{\frac{\hbar}{i}\frac{\partial\Psi}{\partial t}-\frac{\hbar^2}{2m}
\nabla^2\Psi+U\Psi\}d\tau=0
\end{equation}
For that equation to be identically satisfied its integral
expression  must be equal to 0:
\begin{equation}
\label{14} \hbar i\frac{\partial\Psi}{\partial
t}=-\frac{\hbar^2}{2m}\nabla^2\Psi+U\Psi
\end{equation}

which yields the time-dependent Schroedinger
equation.\\

 There is one unpleasant thing about generality of
 equation (\ref{14}). It has been obtained under restriction (\ref{11}) imposed on
 $|\Psi|^2$, that is the requirement that in  (\ref{8}) the
 amplitude $d(\vec{x},t)$ must be a function of the coordinates
 only. On the other hand, from the time-dependent Schroedinger equation
 (\ref{14})
 follows a more general continuity equation:
 \begin{equation}
 \label{15}
\frac{\partial}{\partial
t}|\Psi|^2=\frac{\hbar}{2im}\nabla\cdot[|\Psi|^2 \nabla
(Ln\frac{\Psi^*}{\Psi})]
\end{equation}

Thus the wave equation (\ref{14}) assumes, so-to-speak, its own
life and transcends the restriction imposed on function $\Psi$ in
the process of arriving at this equation.\\

The complex-valued character of the modified action $S$ allows one
to perform an inverse problem: to arrive at the Hamilton-Jacobi
equation by departing from the respective Schroedinger equation
without the usual requirement $\hbar\rightarrow 0$. Let us assume
that in the classical Hamilton-Jacobi (\ref{1}) equation the term
$(\nabla S)^2$ is to be replaced by $(\nabla S)\cdot(\nabla S)^*$
( as we did in the above calculations). We apply the quantum
averaging to the Schroedinger equation (\ref{14}), assume that at
the boundary surface $\Sigma$ of the region of integration
$\Psi_{\Sigma}=0$, and use the Schroedinger substitution
(\ref{add}). As a result, we arrive at the exact averaged (with
the weight $\Psi\Psi^*$) classical Hamilton-Jacobi equation:
\begin{eqnarray}
\label{15a} \int\Psi^*[\frac{\hbar} {i}\frac{\partial}{\partial
t}-\frac{\hbar^2}{2m}\nabla^2+U]\Psi d\Omega\equiv
-\frac{\hbar^2}{2m}\int\Psi^*\nabla\Psi\cdot
\vec{d\Sigma}+\nonumber \\\int[\frac{\hbar}
{i}\Psi^*\frac{\partial\Psi}{\partial
t}+\frac{\hbar^2}{2m}\nabla\Psi\cdot\nabla\Psi^*+\Psi^*U\Psi]d\Omega\equiv\nonumber
\\ \int\Psi^*[\frac{\partial}{\partial t}(\frac{\hbar}{i}Ln \Psi)+
\frac{1}{2m}\nabla (\frac{\hbar}{i}Ln \Psi)\cdot\nabla
(\frac{\hbar}{i}Ln \Psi)^*+U]\Psi d\Omega \equiv\nonumber \\
\int\Psi^*[\frac{\partial S}{\partial t}+\frac{1}{2m}(\nabla
S)\cdot(\nabla S)^* +U]\Psi d\Omega= 0
\end{eqnarray}

If we introduce an average quantity according to the following
$$<...>=\int\Psi^*(...)\Psi d\Omega$$

then (\ref{15a}) yields the classical Hamilton-Jacobi equation (in
the complex domain) with respect to these averaged quantities
\begin{equation}
\label{15b} <\frac{\partial S}{\partial t}>+\frac{1}{2m}<|\nabla
S|^2> +<U>=0
\end{equation}

If, in addition, $$|\Psi^2|\neq f(t)$$ then (\ref{15b}) yields:
\begin{equation}
\label{15c} \frac{<\partial S>}{\partial t}+\frac{1}{2m}<|\nabla
S|^2> +<U>=0
\end{equation}
Interestingly enough, we arrive at the averaged Hamilton-Jacobi
equation (\ref{15b}) or (\ref{15c}) without resorting to the
conventional
 limit $\hbar\rightarrow 0$. Moreover, it becomes clear that function $\Psi$ acquires an
 additional meaning of some "averaging" function, this time in the complex region.\\

We extend the above procedures leading to the Schroedinger
equation to a non-relativistic Hamilton-Jacobi equation for a
particle in an electro-magnetic field:
\begin{equation}
\label{16} \frac{\partial S}{\partial t}+\frac{1}{2m}(\nabla
S-e\vec{A})^2+e\phi=0
\end{equation}
where $e$ is the particle charge, $\phi$ is a scalar potential,
$\vec{A}$ is the vector potential and we set the speed of light
$c=1$. Introducing the Schroedinger substitution (\ref{a1}) and
replacing in (\ref{16}) the second term by the product of the
respective factors expressed in terms of $\Psi$ and its
complex-conjugate $\Psi^*$ we get
\begin{eqnarray}
\label{17} \frac{\hbar}{i}\Psi^*\frac{\partial\Psi}{\partial
t}+\frac{1}{2m}(\hbar^2\nabla\Psi\cdot\nabla\Psi^*+\frac{\hbar}{i}e\Psi\vec{A}\cdot\nabla\Psi^*-
\frac{\hbar}{i}e\Psi^*\vec{A}\cdot\nabla\Psi+e^2A^2\Psi\Psi^*)\nonumber\\
+e\phi\Psi\Psi^*=0
\end{eqnarray}
Integrating (\ref{17}) over a volume $\tau$ and assuming that at
the boundary surface $\Sigma$ of this volume $\Psi|_{\Sigma} =0$
we arrive at the following:
\begin{equation}
\label{i1} \int\Psi^*[\frac{h}{i}\frac{\partial\Psi}{\partial
t}+\frac{1}{2m}(\frac{\hbar}{i}\nabla-e\vec{A})^2\Psi+e\phi\Psi]d\tau=0
\end{equation}

This equation is identically satisfied, if its integral function
is equal to 0. This yields the time-dependent Schroedinger
equation for a non-relativistic spinless particle in an
electro-magnetic field:
\begin{equation}
\label{18} i\hbar\frac{\partial\Psi}{\partial
t}=\frac{1}{2m}(\frac{\hbar}{i}\nabla-e\vec{A})^2\Psi+e\phi\Psi
\end{equation}\\

We  also use this method to obtain the Schroedinger equation for a
system of $j=1,2,...N$  particles. In fact, the respective
Hamilton-Jacobi equation is
\begin{equation}
\label{19} \frac{\partial S}{\partial
t}+\frac{1}{2}\sum_{j=1}^{N}\frac{1}{m_j}\frac{\partial
S}{\partial\vec{x}_j}\cdot\frac{\partial
S}{\partial\vec{x}_j}+U(\vec{x}_k,\vec{x}_n,...)=0
\end{equation}
If we substitute in (\ref{19}) expression (\ref{a1}), require that
all the terms in the sum must be real-valued, and use to this end
the complex-conjugate function $\Psi^*$ we then obtain
\begin{equation}
\label{20} \frac{\hbar}{i}\Psi^*\frac{\partial\Psi}{\partial
t}+\frac{\hbar^2}{2}\sum_{j=1}^{N}\frac{1}{m_j}\frac{\partial
\Psi}{\partial\vec{x}_j}\cdot\frac{\partial
\Psi^*}{\partial\vec{x}_j}+U(\vec{x}_j,\vec{x}_n,...)\Psi\Psi^*=0
\end{equation}
Integrating (\ref{20}) over a hyper-volume
$d\tau=\Pi_{i=1}^Nd\tau_i$ and assuming that at each respective
boundary hyper-surface $\Sigma_i$ function $\Psi_{\Sigma_i}=0$ we
arrive at the following equation
\begin{equation}
\label{21} \int\Psi^*[\frac{\hbar}{i}\frac{\partial\Psi}{\partial
t}-\frac{\hbar^2}{2}\sum_{j=1}^N\frac{1}{m_j}(\frac{\partial}{\partial\vec{x}_j}\cdot\frac{\partial}
{\partial\vec{x}_j})\Psi +U\Psi]d\tau=0
\end{equation}
This  results in the Schroedinger equation for $N$ particles
\begin{equation}
\label{22} i\hbar\frac{\partial\Psi}{\partial
t}=-\frac{\hbar^2}{2}\sum_{j=1}^N\frac{1}{m_i}(\frac{\partial}{\partial\vec{x}_j}\cdot\frac{\partial}
{\partial\vec{x}_j})\Psi +U\Psi\\
\end{equation}\\

In a similar fashion we can obtain Klein-Gordon equation for a
relativistic particle with spin 0 from the relativistic
Hamilton-Jacobi equation for a charged particle in an
electro-magnetic field:
\begin{equation}
\label{23} g^{jk}(\frac{\partial S}{\partial
x^j}+eA_j)\cdot(\frac{\partial S}{\partial x^k}+eA_k)=m^2
\end{equation}
where  $g^{jk}=(1,-1,-1,-1),~j=0,1,2,3,~A_0=\phi,$
$A_{\alpha}(\alpha=1,2,3)=-\vec{A}$ ,we set $c=1$ and the
summation over the repeated indices is adopted.\\

We use the Schroedinger substitution (\ref{a1}) and replace the
right-hand side of (\ref{23}) by the product of the expressions
containing $\Psi$ and its complex conjugate (to be consistent with
the real-valuedness of the left-hand side):
\begin{equation}
\label{24} g^{jk}(\frac{\hbar}{i}\frac{\partial\Psi}{\partial
x^j}+eA_j\Psi)(-\frac{\hbar}{i}\frac{\partial\Psi^*}{\partial
x^k}+eA_k\Psi^*)-m\Psi^*m\Psi=0
\end{equation}
Now we integrate (\ref{24}) over a $4$-volume  $d\Omega
=dx^0dx^1dx^2dx^3$ and assume that function $\Psi|_{\Sigma_j}=0$
at the boundary hyper-surface $\Sigma_j$ of the four-volume. As a
result, (after integrating the first term by parts) we obtain
\begin{equation}
\label{25} \int\Psi^*\{g^{jk}[-\hbar^2
\frac{\partial^2\Psi}{\partial x^j\partial
x^k}+e\frac{\hbar}{i}(A_k\frac{\partial\Psi}{\partial
x^j}+\frac{\partial }{\partial
x^k}A_j\Psi)+e^2A_jA_k\Psi]-m^2\Psi\} d\Omega=0
\end{equation}
This equation is identically satisfied if
\begin{equation}
\label{26} g^{jk}(\frac{\hbar}{i}\frac{\partial}{\partial
x^j}+eA_j)(\frac{\hbar}{i}\frac{\partial}{\partial
x^k}+eA_k)\Psi-m^2\Psi=0
\end{equation}
yielding the Klein-Gordon equation for a zero spin particle.\\

Finally, departing from the relativistic Hamilton-Jacobi equation
(\ref{23}) which we write as follows:
\begin{equation}
\label{27} (\frac{\partial S}{\partial t}+e\phi)^2={(\nabla
S-e\vec{A})^2+m^2}
\end{equation}
we arrive at the Dirac equation. Using the Schroedinger
substitution (\ref{add}) in (\ref{27}) and the fact that the both
sides of the resulting equation containing the complex-valued
function $\Psi$ must be real-valued, we get
\begin{equation}
\label{28} (i\hbar\frac{\partial\Psi}{\partial
t}-e\phi\Psi)^*(i\hbar\frac{\partial\Psi}{\partial
t}-e\phi\Psi)=(\frac{\hbar}{i}\nabla\Psi-e\vec{A}\Psi)^*
(\frac{\hbar}{i}\nabla\Psi-e\vec{A}\Psi)+m^2\Psi\Psi^*
\end{equation}
The right-hand side of (\ref{28}) can be represented as a product
of two factors, if we replace the complex-valued function $\Psi$
with a four-vector function $\hat{\Psi}$ (a bispinor), use Dirac's
matrices $\alpha_j(j=1,2,3)$ and $\beta$\,

\begin{math}
\alpha_1=\bordermatrix{&&\cr
&0&\sigma_x\cr
        &\sigma_x&0\cr};~~~
\alpha_2=\bordermatrix{&&\cr &0&\sigma_y\cr
        &\sigma_y&0\cr};~~~
        \alpha_3=\bordermatrix{&&\cr &0&\sigma_z\cr
         &\sigma_y&0\cr};~~
        \beta=\bordermatrix{&&\cr &1&0\cr
        &0&-1\cr}
                \end{math}\\

and instead of complex-conjugation use Hermite conjugation.\ As a
result, we arrive at the following equation:
\begin{eqnarray}
\label{29} (i\hbar\frac{\partial\hat{\Psi}}{\partial
t}-e\phi\hat{\Psi})^{\dagger}(i\hbar\frac{\partial\hat{\Psi}}{\partial
t}-e\phi\hat{\Psi})&=&\nonumber\\
\{\vec{\alpha}\cdot(\frac{\hbar}{i}\vec{\nabla}-e\vec{A})\hat{\Psi}+
m\beta\hat{\Psi}\}^{\dagger}
\{\vec{\alpha}\cdot(\frac{\hbar}{i}\vec{\nabla}-e\vec{A})\hat{\Psi}
+m\beta\hat{\Psi}\}
\end{eqnarray}
From (\ref{29}) immediately follows the Dirac equation:
\begin{equation}
\label{30} i\hbar\frac{\partial\hat{\Psi}}{\partial
t}=[\alpha\cdot(\frac{\hbar}{i}\vec{\nabla}-e\vec{A})+e\phi+m\beta]\hat{\Psi}
\end{equation}\\
\section{Dissipative Perturbations in Hamilton-Jacobi Equation
lead to the Schroedinger Equation}

As was indicated in the Introduction, 't Hooft \cite{GH} proposed
to permit at the classical level the information loss (via some
dissipative mechanism). This would make possible to get bounded
hamiltonian in quantum formalism and to obtain an apparent
quantization of the orbits which looks very similar to the quantum
structure seen in the real world. As a specific illustration of
the proposal, the work \cite{MB} demonstrated that for a
particular case of two classical damped oscillators the
dissipative term is responsible for a zero point energy "in the
quantum spectrum of the $1$-D linear harmonic oscillator".\\

Importantly, 't Hooft wrote that "making information dissipate is
not easy in continuum theories. It may well be that discrete
degrees of freedom must be added." Here we show how in a $more
~general$ scheme of things (not restricted to some special cases)
adding a dissipative term to the classical equation of motion will
naturally lead to the wave equations, ranging from the Shroedinger
to Klein-Gordon equations.\\

We begin with the second law of Newton for a single particle
written in the following form (e.g.,\cite{AG})
\begin{equation}
\label{30}
 \frac{\partial \vec{p}}{\partial t} +
(\frac{\vec{p}}{m}\cdot\nabla)\vec{p}= \vec{F}(\vec{p},\vec{x},t)
\end{equation}
where the force $\vec{F}$ is a sum of the conservative part
($-\nabla U$) and a dissipative part. Since the left hand side of
(\ref{30}) looks like the respective fluid-dynamical equation, the
information loss argued by 't Hooft is introduced by adding the
perturbation term analogous to the viscous friction term in the
Navier-Stokes equation

\begin{equation}
\label{al2} \nu^q\nabla^2\vec{p} \end{equation}

Here constant $\nu^q$ (to be determined below) can be considered
as some sort of
 "quasi $ kinematic ~~viscosity$". Thus
\begin{equation}
\label{31} \vec{F}=-\nabla U+\nu^q\nabla^2\vec{p}
\end{equation}\\

Inserting (\ref{31}) in (\ref{30}) we obtain
\begin{equation}
\label{32} \frac{\partial \vec{p}}{\partial t} +
(\frac{\vec{p}}{m}\cdot\nabla)\vec{p}=-\nabla
U+\nu^q\nabla^2\vec{p}
\end{equation}
Applying $curl$ to both sides of (\ref{32}) results in the
following:
\begin{equation}
\label{33} \frac{\partial}{\partial
t}\nabla\times\vec{p}-\frac{1}{m}\nabla\times[\vec{p}\times(\nabla\times\vec{p})]
-\nu^q\nabla\times[\nabla\times(\nabla\times\vec{p})]=0
\end{equation}
Equation (\ref{33}) is identically satisfied if
$\nabla\times\vec{p}=0$, or equivalently
\begin{equation}
\label{34} \vec{p}=\nabla S^q
\end{equation}
where $S^q$ is a new "generalized action".\\

Inserting (\ref{34}) in (\ref{32}) and performing some elementary
vector operations we obtain
\begin{equation}
\label{35} \nabla\{\frac{\partial S^q}{\partial
t}+\frac{1}{2m}(\nabla S^q)^2+U-\nu^q\nabla^2 S^q\}=0
\end{equation}
This equation os identically satisfied if $$\frac{\partial
S^q}{\partial t}+\frac{1}{2m}(\nabla S^q)^2+U-\nu^q\nabla^2
S^q=0$$ yielding the Hamilton-Jacobi equation with dissipation,
represented by small perturbation term $\nu^q\nabla^2 S^q$.\\

We use the Schroedinger substitution (\ref{a1}) in (\ref{35})
(which transforms it into a nonlinear homogenous of order $2$
partial differential equation) and get
\begin{equation}
\label{36} -i\hbar\Psi\frac{\partial\Psi}{\partial
t}-(\nabla\Psi)^2(\frac{\hbar^2}{2m}-\frac{\nu^q\hbar}{i})-
\nu^q\frac{\hbar}{i}\Psi\nabla^2\Psi+U\Psi^2=0
\end{equation}
Since the experiments demonstrate that at the microlevel (at least
in a majority of cases) the superposition principle holds, the
equation which should follow from (\ref{36}) must be linear. This
determines the $"quasi~kinematic~viscosity"$ $\nu^q$:
$$\nu^q=\frac{i\hbar}{2m}$$
Upon substitution of this expression in (\ref{36}) we arrive at
the Schroedinger equation
\begin{equation}
\label{37} i\hbar\frac{\partial\Psi}{\partial
t}=-\frac{\hbar^2}{2m}\nabla^2\Psi+U\Psi
\end{equation}

Interestingly enough, the introduction of the information loss (in
a form of small perturbations)\footnote{Physically this smallness
is determined by a comparison on a dimensional basis of the
viscous term $\hbar p/mL^2\sim m\lambda^2/LT^2$ (where $L$ is the
characteristic length and $\lambda$ is the wavelength) and the
dynamic term $p^2/mL\sim mL^2/T$. Their ratio $\lambda^2/L^2$ is
tiny, when we are dealing with classical phenomena. This is
compatible with the view of considering a classical path as a
geometrical optics limit $\lambda\rightarrow 0$ of the wave
propagation.} is compatible with $fractalization$ of the
deterministically defined classical path (one-dimensional curve)
which  degenerates into a quantum path, whose Hausdorff
dimension is $2$ \cite{LA,AG2}.\\

It has also turned out that by introducing the dissipative term
(small perturbations) into the Hamilton-Jacobi equation (\ref{16})
for a charged particle in an electro-magnetic field we arrive at
the respective Schroedinger equation. This is done by keeping in
mind that now the generalized momentum is $\nabla S-e\vec{A}$ and
not simply $\nabla S$:
\begin{equation}
\label{38} \frac{\partial S}{\partial t}+\frac{1}{2m}(\nabla
S-e\vec{A})^2+e\phi=\nu^q\nabla\cdot(\nabla S-e\vec{A})
\end{equation}
Using the Schroedinger substitution (\ref{a1}) in (\ref{38}) and
performing some elementary vector operations we arrive at the
following
\begin{eqnarray}
\label{39} \Psi\{-i\hbar\frac{\partial\Psi}{\partial
t}+\frac{1}{2m}(e^2A^2\Psi-2\frac{\hbar}{i}e\vec{A}\cdot\nabla\Psi)+e\phi\Psi-\nonumber\\
\nu^q(\frac{\hbar}{i}\nabla^2\Psi-e\Psi\nabla\cdot\vec{A})\}+
(\nabla\Psi)^2(\nu^q\frac{\hbar}{i}-\frac{\hbar^2}{2m})=0
\end{eqnarray}
By requiring this equation to be linear we get the following value
of constant $\nu^q$
$$\nu^q=i\frac{\hbar}{2m}$$
 Inserting this back in
(\ref{39}) we arrive at the respective Schroedinger equation:
\begin{equation}
\label{40} i\hbar\frac{\partial\Psi}{\partial
t}=\frac{1}{2m}(\frac{\hbar}{i}\nabla-e\vec{A})^2\Psi+e\phi\Psi
\end{equation}\\
Interestingly enough, the same idea enables us to  derive  the
Klein-Gordon equation for a charged relativistic particle of spin
$0$ in an electro-magnetic field. To this end we add a small
perturbation term $$\nu^q g^{jk}\frac{\partial}{\partial
x^k}(\frac{\partial S}{\partial x^j}+eA_j)$$ (where $\nu^q$ is to
be determined) to the right hand side of (\ref{26}), use the
Schroedinger substitution and get
\begin{eqnarray}
\label{41} g^{jk}(\frac{\hbar}{i}\frac{\partial\Psi}{\partial
x^j}+e\Psi A_j)(\frac{\hbar}{i}\frac{\partial\Psi}{\partial
x^k}+e\Psi A_k)=m^2\Psi^2+\nonumber\\
\nu^qg^{jk}(\frac{\partial^2\Psi}{\partial x^j\partial
x^k}-\frac{\hbar}{i}\frac{\partial\Psi}{\partial
x^j}\frac{\partial\Psi}{\partial x^k}+e\Psi\frac{\partial
A_j}{\partial x^k})
\end{eqnarray}
Linearity requirement imposed on this equation determines the
value of constant $\nu^q$: $$\nu^q=i\hbar$$ Inserting this value
back in (\ref{41}) and performing some elementary calculations we
arrive at the Klein-Gordon equation for a charged relativistic
particle of spin $0$ in an electro-magnetic field:
\begin{equation}
\label{42}
 g^{jk}(\frac{\hbar}{i}\frac{\partial}{\partial
x^j}+eA_j)(\frac{\hbar}{i}\frac{\partial}{\partial
x^k}+eA_k)\Psi=m^2\Psi
\end{equation}\\

Since this idea clearly works for particles with zero spin, it is
naturally to ask whether it would work for particles with a spin.
Her one must be a little bit more ingenious in choosing the
appropriate dissipative term to be introduced into the
Hamilton-Jacobi equation. If we consider a classical  charged
particle in the electro-magnetic field it has an additional energy
$U=-\vec{\mu}\cdot\vec{H}$ due to an interaction of the magnetic
moment $\vec{\mu}$ and the magnetic field $\vec{H}$.\\

In terms of the vector potential this energy is
\begin{equation}
\label{43} U=-\vec{\mu}\cdot(\nabla\times e\vec{A})\equiv
div(\vec{\mu}\times\vec{A})
\end{equation}

Experiments demonstrated that the magnetic moment $\vec{\mu_e}$ of
an electron is proportional to its spin $\vec{s}$:
\begin{equation}
\label{44} \vec{\mu_e}=\frac{\hbar}{m}\vec{s}
\end{equation}
It is remarkable that the coefficient of proportionality in
(\ref{44}) has the dimension of kinematic viscosity! Its magnitude
is twice the magnitude of the $"quasi~kinematic~viscosity~ \nu^q$.
Let us denote the new coefficient by $\nu^s$. If we substitute
(\ref{44}) in (\ref{43}) we obtain
\begin{equation}
\label{45} U=-\nu^s\nabla\cdot(\vec{s}\times\vec{A})
\end{equation}
This expression has a structure of the dissipative term introduced
earlier ( see \ref{al2}) in the Hamilton-Jacobi equation
(\ref{38}). Therefore we rewrite this equation with the additional
"dissipative" term (\ref{45})
\begin{equation}
\label{46} \frac{\partial S}{\partial t}+\frac{1}{2m}(\nabla
S-e\vec{A})^2+e\phi=\nabla\cdot[\nu^q(\nabla
S-e\vec{A})-\nu^s(\vec{s}\times\vec{A})]
\end{equation}\

We use Schroedinger substitution (\ref{a1}) in (\ref{46}), perform
some vector operations and obtain
\begin{eqnarray}
\label{47} -i\hbar\Psi\frac{\partial\Psi}{\partial
t}+\frac{1}{2m}(\frac{\hbar}{i}\nabla\Psi-e\vec{A}\Psi)^2+e\phi\Psi^2=\nonumber\\
\Psi^2\frac{\hbar}{i}\nu^q\nabla\cdot(\frac{\nabla\Psi}{\Psi})-
\Psi^2\nu^s\nabla\cdot(\vec{s}\times\vec{A})
\end{eqnarray}
Eliminating nonlinearity (which uniquely defines $\nu^q$ as
$i\hbar/2m$) and taking into account that function $\Psi$ now
depends on the $z$-component of the spin $\vec{s}$ (that is, it
becomes a $2\times 1$ vector-column function) we have to replace
vector $\vec{s}$ by the Pauli matrices $\hat{\vec{\sigma}}$. As a
result, we arrive at the Pauli equation:
\begin{equation}
\label{48} i\hbar\frac{\partial\Psi}{\partial
t}=\frac{1}{2m}(\frac{\hbar}{i}\nabla-e\vec{A})^2\Psi+e\phi\Psi-
\frac{e\hbar}{m}(\hat{\vec{\sigma}}\cdot\vec{H})
\end{equation}\\

Since the  method of information loss introduced into the
classical Hamilton-Jacobi equations has turned out to be
successful, we apply it to the simple case of a particle in the
gravitational field. The Hamilton-Jacobi equation in this case is
\begin{equation}
\label{49} g^{jk}S_{;j}S_{;k}-m^2=0
\end{equation}
where $g^{jk}$ is the metric tensor, $j,k=0,1,2,3$, $semicolon$
denotes covariant differentiation, and we set $c=1$.\

Now I add to the right-hand side of (\ref{49}) the dissipative
term in the form used in the above calculations, that is
$div(\nu^q\nabla S)$. However, this time, instead of the
conventional derivatives, I use the covariant derivatives and
replace the constant scalar $\nu^q$ by a tensor function
$\nu^{jk}$. As a result, equation (\ref{49}) becomes:
\begin{equation}
\label{50} g^{jk}S_{;j}S_{;k}-m^2=(\nu^{jk}S_{;k})_{;j}
\end{equation}\

By using the Schroedinger substitution (\ref{a1}) in (\ref{50})
and performing some standard calculations we obtain the following
\begin{eqnarray}
\label{51} -\hbar^2g^{jk}\frac{\partial\Psi}{\partial
x^j}\frac{\partial\Psi}{\partial
x^k}-m^2\Psi^2+\frac{\hbar}{i}\nu^{jk}\frac{\partial\Psi}{\partial
x^j}\frac{\partial\Psi}{\partial
x^k}-\frac{\hbar}{i}\nu^{jk}_{;j}\Psi\frac{\partial\Psi}{\partial
x^k}\nonumber\\
-\frac{\hbar}{i}\nu^{jk}\Psi(\frac{\partial^2\Psi}{\partial
x^j\partial x^k}+\Gamma_{kj}^n\frac{\partial\Psi}{\partial
x^n})-m^2\Psi^2=0
\end{eqnarray}
where $\Gamma_{jk}^n$ is the Ricci tensor. We require this
equation to be linear, which uniquely determines the value of the
tensor $\nu^{jk}$:
\begin{equation}
\label{52} \nu^{jk}=i\hbar g^{jk}
\end{equation}
Since $g^{jk}_{;j}\equiv 0$ equation (\ref{51}) yields
\begin{equation}
\label{53} g^{jk}\frac{\partial^2\Psi}{\partial x^j\partial
x^k}-\frac{1}{\sqrt{-g}}\frac{\partial}{\partial
x^l}(\sqrt{-g}g^{nl})\frac{\partial\Psi}{\partial
x^n}+\kappa^2\Psi=0
\end{equation}
where $\kappa=m/\hbar$.\\

As a particular example we consider the centrally symmetric
gravitational field with the Schwarzchild metric:
\begin{eqnarray}
\label{54}
g^{jk}=0,j\neq k;~~~g^{00}=\frac{1}{1-r_g/r};~~~g^{11}=-(1-\frac{r_g}{r});\nonumber\\
g^{22}=-\frac{1}{r^2};~~~g^{33}=-(1-\frac{r_g}{r});~~~g=|g^{jk}|
=-\frac{1}{r^4\sin^2\theta};~~~r_g=2mG
\end{eqnarray}\
Equation (\ref{53}) is then
\begin{eqnarray}
\label{55} \frac{1}{1-r_g/r}\frac{\partial^2\Psi}{\partial
t^2}-(1-\frac{r_g}{r})\frac{\partial^2\Psi}{\partial r^2}-
\frac{1}{r^2
sin^2\theta}\frac{\partial^2\Psi}{\partial\phi^2}-\nonumber\\
\frac{1}{r^2}\frac{\partial^2\Psi}{\partial\theta^2}-\frac{2}{r}(1-\frac{3}{2}\frac{r_g}{r})
\frac{\partial\Psi}{\partial
r}-\frac{1}{r^2}cot\theta\frac{\partial\Psi}{\partial\theta}+\kappa^2\Psi=0
\end{eqnarray}

\section{Conclusion}

Physical phenomena can only be described as either particle-like
or wave-like phenomena. Consequently, the critical question
arises: Does the complex-valued wave function $\Psi$ represent
reality, or is it only an intricate device to deal with something
we don't have a complete knowledge of? \\

Bohm \cite{B} proposed to remove such indeterminacy and thus to
answer the above question by introducing hidden variables right
into the quantum-mechanical equations. This idea was shelved for
quite some time until recently, when it was given a new lease on
life by 't Hooft \cite{GH} who treated the hidden variables
differently. He proposed to establish the physical link between
classical and quantum world by starting from the continuum
equations of classical mechanics and include into them a specially
chosen dissipative function.\\

We explore this idea, and it has demonstrated its effectiveness.
Such an approach has allowed us to elucidate the indeterminacy
inherent in quantum-mechanical description. It has turned out that
the latter is due to the information loss. An appropriate analogy
would be the loss of knowledge of an initial velocity of a solid
body falling in a viscous medium, if we know only its terminal
velocity.\\

As a result, the wave-like quantum mechanics turns out to follow
from the particle-like classical mechanics due to embedding in the
latter a dissipation "device" responsible for the loss of
information. Indeed, the initial precise information about the
classical trajectory of a particle is lost in quantum mechanics
due to the "dissipative spread" of the trajectory and its
transformation into a fractal path with  Hausdorff dimension of
$2$ in a simple case of a spinless particle
\cite{LA},\cite{AG2}.\\

The approach suggested by 't Hooft and realized (in a very general
way) in this paper has not only made it possible to elementary
derive the basic quantum-mechanical equations, instead of
traditionally postulating them, but also seems to be applicable to
more complex and intriguing problems, as for example,
a relativistic particle in the gravitational field.\\

{\bf{Acknowledgments}} The author would like to thank
Prof.V.Granik for the illuminating discussions.

\end{document}